\documentclass[11pt]{article}

\usepackage[numbers]{natbib}
\usepackage[margin=1in]{geometry}
\usepackage{hyperref}
\usepackage{tabularx}
\usepackage{ragged2e}
\newcolumntype{Y}{>{\RaggedRight\arraybackslash}X}
\usepackage{booktabs} % For formal tables
\usepackage{nicefrac}       % compact symbols for 1/2, etc.
\usepackage{eurosym}
\usepackage{graphicx}

\newcommand{\ra}[1]{\renewcommand{\arraystretch}{#1}}

\begin{document}

\title{Mitigating Bias in Algorithmic Hiring: Evaluating Claims and Practices}

\author{Manish Raghavan\thanks{Cornell University} \and Solon
Barocas\thanks{Microsoft Research and Cornell University} \and Jon Kleinberg\footnotemark[1] \and Karen Levy\footnotemark[1]}
\date{}

\maketitle
\begin{abstract}
There has been rapidly growing interest in the use of algorithms in hiring, especially as a means to address or mitigate bias. Yet, to date, little is known about how these methods are used in practice.
How are algorithmic assessments built, validated, and examined for bias?
In this work, we document and analyze the claims and practices of companies offering algorithms for employment assessment.
In particular, we identify vendors of algorithmic pre-employment assessments (i.e., algorithms to screen candidates), document what they have disclosed about their development and validation procedures, and evaluate their practices, focusing particularly on efforts to detect and mitigate bias.
Our analysis considers both technical and legal perspectives.
Technically, we consider the various choices vendors make regarding data collection and prediction targets, and explore the risks and trade-offs that these choices pose.
We also discuss how algorithmic de-biasing techniques interface with, and create challenges for, antidiscrimination law.
\end{abstract}

\section{Introduction}
The study of algorithmic bias and fairness in machine learning has quickly matured into a field of study in its own right, delivering a wide range of formal definitions and quantitative metrics.
As industry takes up these tools and accompanying terminology, promises of eliminating algorithmic bias using computational methods have begun to proliferate. In some cases, however, rather than forcing precision and specificity, the existence of formal definitions and metrics has had the paradoxical result of giving undue credence to vague claims about ``de-biasing'' and ``fairness.''

In this work, we use  algorithmic pre-employment assessment as a case study to show how formal definitions of fairness allow us to ask focused questions about the meaning of ``fair'' and ``unbiased'' models.
The hiring domain makes for an effective case study because of both its prevalence and its long history of bias.
We know from decades of audit studies that employers tend to discriminate against women and ethnic minorities~\cite{bertrand2004emily,bendick1997employment,bendick2012developing,johnson2016if}, and a recent meta-analysis suggests that little has improved over the past 25 years~\cite{quillian2017meta}.
Citing evidence that algorithms may help reduce human biases~\cite{houser2019can,kleinberg2017human}, advocates argue for the adoption of algorithmic techniques in hiring~\cite{chamorro2019should,cowgill2018bias}, with a variety of computational metrics proposed to identify and prevent unfair behavior~\cite{feldman2015certifying}. But to date, little is known about how these methods are used in practice.

One of the biggest obstacles to empirically characterizing industry practices is the lack of publicly available information.
Much technical work has focused on using computational notions of equity and
fairness to evaluate specific models or
datasets~\cite{angwin2016machine,buolamwini2018gender}. Indeed, when these
models are available, we can and should investigate them to identify potential points of concern. But what do we do when we have little or no access to models or the
data that they produce? Certain models may be completely inaccessible to the public, whether for practical or legal reasons, and attempts
to audit these models by examining their training data or outputs might
place users' privacy at risk.
With algorithmic pre-employment assessments, we find that this is very much the case: models, much less the sensitive employee data used to construct them, are in general kept private.
As such, the only information we can consistently glean about industry practices is limited to what companies publicly disclose.
Despite this, one of the key findings of our work is that even without
access to models or data, we can still learn a considerable amount by investigating
what corporations disclose about their practices for developing, validating, and
removing bias from these tools.

\paragraph*{Documenting claims and evaluating practices.}
Following a review of firms offering recruitment technologies, we identify 18 vendors of pre-employment assessments. We document what each company has disclosed about its practices and consider the implications of these claims.
In so doing, we develop an understanding of industry attempts to address bias
and what critical issues have been left unaddressed.

Prior work has sought to taxonomize the points at which bias can enter machine
learning systems, noting that the choice of target variable or outcome to
predict, the training data used, and labelling of examples are all potential
sources of disparities~\cite{barocas2016big,kleinberg2019discrimination}.
Following these frameworks, we seek to understand how practitioners handle these key decisions in the machine learning pipeline.
In particular, we surface choices and trade-offs vendors face with regard to the collection of data, the ability to validate on representative populations, and the effects of discrimination law on efforts to prevent bias.
The heterogeneity we observe in vendors' practices indicates evolving industry norms that are sensitive to concerns of bias but lack clear guidance on how to respond to these worries.

Of course, analyzing publicly available information has its limitations. We are
unable, for example, to identify issues that any particular model might raise in
practice. Nor can we be sure that vendors aren't doing more behind the scenes to
ensure that their models are non-discriminatory. And while other publicly
accessible information (e.g., news articles and videos from conferences) might
offer further details about vendors' practices, for the sake of consistent
comparison, we limit ourselves to statements on vendors' websites. As such, our
analysis should not be viewed as exhaustive; however, as we will see, it is
still possible to draw meaningful conclusions and characterize industry trends
through our methods. One notable limitation we encounter is the lack of
information about the validity of these assessments. It is of paramount
importance to know the extent to which these tools actually work, but we cannot
do so without additional transparency from vendors.

We stress that our analysis is not intended as an expos{\'e} of
industry practices. Many of the vendors we study exist precisely because they
seek to provide a fairer alternative to traditional hiring practices.
Our hope is that this work
will paint a realistic picture of the landscape of algorithmic techniques in
pre-employment assessment and offer recommendations for their effective and appropriate use.

\paragraph*{Organization of the rest of the paper.}
Section~\ref{sec:background} contains an overview of 
pre-employment assessments, their history, and relevant legal precedents. In
Section~\ref{sec:empirical}, we systematically review vendors of algorithmic screening tools and 
empirically characterize their practices based on the claims that they make. We analyze
these practices in detail in Sections~\ref{sec:analysis} and~\ref{sec:debiasing}
from technical and legal perspectives, examining ambiguities and particular
causes for concern.
We provide concluding
thoughts and recommendations in Section~\ref{sec:discussion}.

\section{Background}
\label{sec:background}
\paragraph*{Pre-employment assessments in the hiring pipeline.}
Hiring decisions are among the most consequential that individuals face,
determining key aspects of their lives, including where they live and how much they
earn. These decisions are similarly impactful for employers, who face
significant financial pressure to make high-quality hires quickly and
efficiently~\cite{mariotti2017talent}. As a result, many employers seek tools
with which to optimize their hiring processes.

Broadly speaking, there are four distinct stages of the hiring pipeline, though
the boundaries between them are not always rigid: sourcing, screening,
interviewing, and selection~\cite{bogen2018help}. Sourcing consists of building
a candidate pool, which is then screened to choose a subset to interview.
Finally, after candidates are interviewed, selected candidates receive offers.
We will focus on \textit{screening}, and in particular,
pre-employment assessments that algorithmically evaluate candidates.
This includes, for example, questionnaires and video interviews that are 
analyzed automatically.

Prior work has considered the
rise of algorithmic tools in the context of hiring, highlighting the concerns that they raise for fairness. Bogen and Rieke
provide an overview of the various ways in which algorithms are being introduced
into this pipeline, with a focus on their implications for equity~\cite{bogen2018help}. Garr surveys a number of platforms designed to
promote diversity and inclusion in hiring~\cite{garr2019diversity}.
S{\'a}nchez-Monedero et al.~\cite{sanchez2019does} analyze
some of the vendors considered here from the perspective of UK law, addressing
concerns over both discrimination and data protection.
Broadly
considering the use of data science in HR-related activities, Cappelli 
et al.~identify several practical challenges to implementation of algorithmic systems in hiring, and propose a framework to help address
them~\cite{cappelli2018artificial}. Ajunwa provides a legal framework to consider the problems algorithmic tools introduce and argues against subjective targets like ``cultural fit''~\cite{ajunwa2019paradox}.
Kim also raises legal concerns over the use of algorithms in hiring in both
advertising and screening contexts~\cite{kim2018big,kim2020manipulating}.

Scholars in the field of Industrial-Organizational (IO) Psychology
have also begun to grapple with the variety of new pre-employment assessment
methods and sources of information enabled by algorithms and big
data~\cite{guzzo2015big}. Chamorro-Prezumic et al. find that academic research
has been unable to keep pace with rapidly evolving technology, allowing vendors
to push the boundaries of assessments without rigorous independent
research~\cite{chamorro2016new}. A 2013 report by the National Research Council summarizes a number of ethical
issues that arise in pre-employment assessment, including the role of human intervention, the provision of feedback to candidates, and the goal of hiring for ``fit,'' especially in light of modern
data sources~\cite{national2013new}. And although proponents argue that pre-employment assessments can push back against human biases~\cite{chamorro2019should}, assessments (especially data-driven algorithmic ones) run the risk of codifying inequalities while providing a veneer of objectivity.

\paragraph*{A history of equity concerns in assessment.}
Pre-employment assessments date back to examinations for
the Chinese civil service thousands of years ago~\cite{haney1982employment}. In
the early 1900's, the idea that assessments could reveal innate cognitive
abilities gained traction in Western industrial and academic circles, leading to
the formation of Industrial Psychology as an academic
discipline~\cite{munsterberg1998psychology,gerhardt1916scientific,kemble1916testing}.
During the two World Wars, the U.S. government turned to these assessments in an
attempt to quantify soldiers' abilities, paving the way for their
widespread adoption in postwar
industry~\cite{baritz1960servants,dubois1970history,dunnette1979personnel}.
Historically, these assessments were primarily behavioral or cognitive in
nature, like the Stanford-Binet IQ test~\cite{terman1916measurement}, the
Myers-Briggs type indicator~\cite{myers1962myers}, and the Big Five personality
traits~\cite{norman1963toward}. IO Psychology
remains a prominent component of these modern assessment tools---many vendors
we examine employ IO psychologists who work with data
scientists to create and validate assessments.

Cognitive assessments have imposed adverse
impacts on minority populations since their introduction into mainstream
use~\cite{tyler1947psychology,ruda1968racial,national1989fairness}.
Critics have long contended that observed group differences in test outcomes indicated flaws in
the tests themselves~\cite{cravens1978triumph}, and a growing consensus has
formed around the idea that while assessments do have some predictive validity,
they often disadvantage minorities despite the fact that minority candidates
have similar real-world job performance to their white
counterparts~\cite{national1989fairness}.\footnote{Disparities in assessment outcomes for minority populations are not limited to
pre-employment assessments. In the education literature, the adverse impact of
assessments on minorities is well-documented~\cite{madaus2001adverse}. This has
led to a decades-long line of literature seeking to measure and mitigate the
observed disparities (see~\cite{hutchinson201950} for a  survey).}

The American Psychological Association (APA) recognizes these concerns as
examples of ``predictive bias'' (when an assessment systematically over- or
under-predicts scores for a particular group) in its Principles for the
Validation and Use of Personnel Selection
Procedures~\cite{society2003principles}. The APA Principles consider several
potential definitions of fairness, and while they encourage practitioners to
identify and mitigate predictive bias, they explicitly reject the view that
fairness requires equal outcomes~\cite{society2003principles}. As we will see,
this focus on predictive bias over outcome-based definitions of fairness forms
interesting connections and contrasts with U.S. employment discrimination law.

\paragraph*{A brief overview of U.S. employment discrimination law.}
Title VII of the Civil Rights Act of 1964 forms the basis of regulatory
oversight regarding discrimination in employment. It prohibits discrimination
with respect to a number of protected attributes (``race, color, religion, sex
and national origin''), establishing the Equal Employment Opportunity Commission
(EEOC) to ensure compliance~\cite{titlevii}. The EEOC, in turn, issued the
Uniform Guidelines on Employment Selection Procedures in 1978 to set standards
for how employers can choose their employees.

According to the Uniform Guidelines~\cite{equal1978department}, the gold standard for pre-employment
assessments is \textit{validity}: the outcome of a test should say something
meaningful about a candidate's potential as an employee. The EEOC accepts three
forms of evidence for validity: criterion, content, and construct. Criterion
validity refers to predictive ability: do test scores correlate with
meaningful job outcomes (e.g., sales numbers)? An assessment with content
validity tests candidates in similar situations to ones that they will encounter
on the job (e.g., a coding interview). Finally, assessments demonstrate construct validity if they test for
some fundamental characteristic (e.g., grit or leadership) required for good job
performance.

When is an assessment legally considered discriminatory? Based on existing
precedent, the Uniform Guidelines provide two avenues to challenge an
assessment: disparate treatment and disparate impact~\cite{barocas2016big}.
Disparate treatment is relatively straightforward---it is illegal to explicitly
treat candidates differently based on categories protected under Title VII~\cite{equal1978department,titlevii}. Disparate impact is more nuanced,
and while we provide an overview of the process here, we refer the reader
to~\cite{barocas2016big} for a more complete discussion.

Under the Uniform Guidelines, the rule of thumb to decide when a disparate
impact case can be brought against an employer is the ``\nicefrac{4}{5} rule'':
if the selection rate for one protected group is less than \nicefrac{4}{5} of
that of the group with the highest selection rate, the employer may be at
risk~\cite{equal1978department}. If a significant disparity in selection rates
is established, an employer may defend itself by showing that its selection
procedures are both valid and necessary from a business
perspective~\cite{equal1978department}. Even when a business necessity has been
established, an employer can be held liable if the plaintiff can produce an
alternative selection procedure with less adverse impact that the employer could
have used instead with little business
cost~\cite{equal1978department}.\footnote{It should be noted that this
  description is based on a particular (although the most common) interpretation
  of Title VII. Some legal scholars contend that Title VII offers stronger
  protections to minorities~\cite{bornstein2018antidiscriminatory,kim2016data},
  and there is disagreement on how (or whether) to operationalize the
  \nicefrac{4}{5} rule through statistical
  tests~\cite{shoben1978differential,cohn1979use,shoben1979defense,cohn1979statistical}.
  In this work, we will not consider alternative interpretations of Title VII,
  nor will we get into the specifics of how exactly to detect violations of the
  \nicefrac{4}{5} rule.} Ultimately, both the APA Principles and the Uniform
  Guidelines agree that validity is fundamental to a good
  assessment.\footnote{Many psychologists disagree with the specific conception
    of validity endorsed by the Uniform
  Guidelines~\cite{mcdaniel2011uniform,salas2011,biddle2008uniform}; however,
there is broad agreement that some form of validation is necessary.} And while
validity can be used as a defense against disparate selection rates, we will see
that the Uniform Guidelines' emphasis on outcome disparities and the
\nicefrac{4}{5} rule significantly impacts vendors' practices.

\section{Empirical Findings}
\label{sec:empirical}
\subsection{Methodology}

\paragraph{Identifying companies offering algorithmic pre-employment assessments.} In order to get a broad overview of the emerging industry surrounding
algorithmic pre-employment assessments, we conducted a systematic review of
assessment vendors with English-language websites. To identify relevant companies, we consulted Crunchbase's list of the top 300 start-ups (by funding amount) under its ``recruiting'' category.\footnote{\url{https://www.crunchbase.com/hub/recruiting-startups}} Originally developed as a platform to track start-ups, Crunchbase now offers information on public and private companies, providing details on funding and other investment activity. While Crunchbase is not an exhaustive list of all companies working in an industry, it is an often-used resource for tracking developments in start-up companies. Companies can create profiles for themselves, subject to validation.\footnote{\url{https://support.crunchbase.com/hc/en-us/articles/115011823988-Create-a-Crunchbase-Profile}} We supplemented this list with an inventory of relevant companies found in recent reports by Upturn~\cite{bogen2018help}, a technology research and advocacy firm focused on civil rights, and RedThread Research~\cite{garr2019diversity}, a research and advisory firm specializing in new technologies for human resource management. This resulted in 22 additional companies, for a combined total of 322. There was substantial overlap between the three sources considered.

Thirty-nine of these companies did not have English-language websites, so we excluded them.
Recall that the hiring pipeline has four primary stages (sourcing, screening, interviewing, and selection); we ruled out vendors that do not provide assessment services at the screening stage, leaving us with 45 vendors.
Note that this excluded companies that merely provide online job boards or marketplaces like Monster.com and Upwork.
Twenty-two of the remaining vendors did not obviously use any predictive technology
(e.g., coding interview platforms that only evaluated correctness or rule-based
screening) or did not offer explicit assessments (e.g., scraping candidate
information from other sources), and an additional 5 did not provide enough
information for us to make concrete determinations, leaving us with 18 vendors in our sample. With these 18 vendors, in April 2019,\footnote{Our empirical findings are specific to this moment in time; practices and documentation may have changed since then.} we recorded administrative information available on
Crunchbase (approximate number of employees, location, and total funding) and undertook a review of their claims and practices, which we explain below.

\paragraph{Documenting vendors' claims and practices.}
Based on prior frameworks intended to interrogate machine learning pipelines for bias~\cite{barocas2016big,kleinberg2019discrimination}, we ask the following questions of vendors:
\begin{itemize}
  \item What types of assessments do they provide (e.g., questions, video interviews, or games)? [Features]
  \item What is the outcome or quality that these assessments aim to predict (e.g., sales revenue, annual review score, or grit)? [Target variable]
  \item What data are used to develop the assessment (e.g., the client's or the vendor's own data)? [Training data]
  \item What information do they provide regarding validation processes (e.g., validation studies or whitepapers)? [Validation]
  \item What claims or guarantees (if any) are made regarding
    bias or fairness? When applicable, how do they achieve these guarantees? [Fairness]
\end{itemize}

To answer these questions, we exhaustively searched the websites of each company.
This included following all internal links, downloading any reports or whitepapers they provided, and watching
webinars found on their websites.
Almost all vendors provided an option to request a demo; we avoided doing so since our focus is on
accessible and public information. Sometimes, company websites were quite sparse
on information, and we were unable to conclusively answer all questions for all
vendors.

\subsection{Findings}
\begin{figure*}[ht]
  \centering
  \includegraphics[width=.75\textwidth]{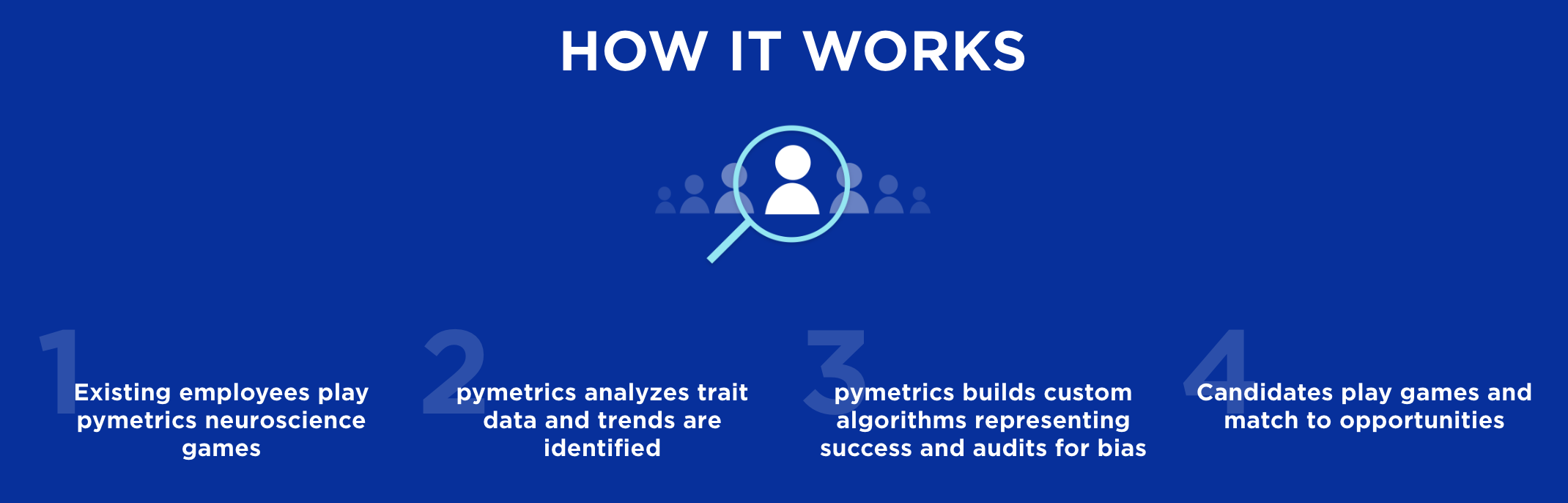}
  \caption{Description of the pymetrics process
  (screenshot from the pymetrics website: \url{https://www.pymetrics.com/employers/})}
  \label{fig:pymetrics}
\end{figure*}

\begin{table*}
  \small
  \centering
\begin{tabular}{lllll}
\toprule
           Vendor name & \shortstack[l]{Assessment types\\\relax[Features]} & \shortstack[l]{Custom? [Target\\\relax\& Training data]} & \shortstack[l]{Validation info\\\relax[Validation]} & \shortstack[l]{Adverse impact\\\relax[Fairness]} \\
\midrule
           8 and Above &                                       phone, video &                                                  S &                                                 -- &                                   bias mentioned \\
              ActiView &                                      VR assessment &                                                  C &                                 validation claimed &                                   bias mentioned \\
 Assessment Innovation &                                   games, questions &                                                 -- &                                                 -- &                                   bias mentioned \\
              Good\&Co &                                          questions &                                               C, P &                                   multiple studies &                                   adverse impact \\
                Harver &                                   games, questions &                                                  S &                                                 -- &                                               -- \\
               HireVue &                            games, questions, video &                                               C, P &                                                 -- &                                         4/5 rule \\
            impress.ai &                                          questions &                                                  S &                                                 -- &                                               -- \\
               Knockri &                                              video &                                                  S &                                                 -- &                                   bias mentioned \\
                  Koru &                                          questions &                                                  S &                                   some description &                                   adverse impact \\
    LaunchPad Recruits &                                   questions, video &                                                 -- &                                                 -- &                                   bias mentioned \\
           myInterview &                                              video &                                                 -- &                                                 -- &                                       compliance \\
               Plum.io &                                   questions, games &                                                  S &                                 validation claimed &                                   bias mentioned \\
        PredictiveHire &                                          questions &                                                  C &                                                 -- &                                         4/5 rule \\
             pymetrics &                                              games &                                                  C &                                   small case study &                                         4/5 rule \\
             Scoutible &                                              games &                                                  C &                                                 -- &                                               -- \\
             Teamscope &                                          questions &                                               S, P &                                                 -- &                                   bias mentioned \\
             ThriveMap &                                          questions &                                                  C &                                                 -- &                                   bias mentioned \\
                  Yobs &                                              video &                                               C, S &                                                 -- &                                   adverse impact \\
\bottomrule
\end{tabular}

\caption{Examining the websites of vendors of algorithmic pre-employment assessments, we answer a number of questions regarding their assessments in relation to questions of fairness and bias. This involves exhaustively searching their websites, downloading whitepapers they provide, and watching webinars they make available. This table presents our findings. The ``Assessment types'' column gives the types of assessments each vendor offers. In the ``Custom?'' column, we consider the source of data used to build an assessment: C denotes ``custom'' (uses employer data), S denotes ``semi-custom''
(qualitatively tailored to employer without data) and P denotes ``pre-built.'' The ``Validation?'' column contains information vendors publicly provided about their validation processes. In the ``Adverse impact'' column, we recorded phrases found on vendors' websites addressing concerns over bias.}
\label{tab:results}
\end{table*}

In our review, we found 18 vendors providing algorithmically driven
pre-employment assessments. Those that had available funding information on
Crunchbase (16 out of 18) ranged in funding from around \$1 million to \$93
million. Most vendors (14) had 50 or fewer employees, and half (9) were based in
the United States. 15 vendors were present in Crunchbase's ``Recruiting
Startups'' list; the
remaining vendors were taken from reports by Upturn~\cite{bogen2018help} and
RedThread Research~\cite{garr2019diversity}. Many vendors were present in all of
these sources. Table~\ref{tab:results} summarizes
our findings. Table~\ref{tab:admin} in Appendix~\ref{app:admin} contains
administrative information about the vendors we included.

\paragraph*{Assessment types.}
The types of assessments offered varied by vendor. The most popular assessment
types were questions (11 vendors), video interview analysis (6 vendors), and
gameplay (e.g., puzzles or video games) (6 vendors). Note that many vendors
offered multiple types of assessments. Question-based assessments included
personality tests, situational judgment tests, and other formats. For video
interviews, candidates were typically either asked to record answers to
particular questions or more free-form ``video resumes'' highlighting their
strengths. These videos are then algorithmically analyzed by vendors.

\paragraph*{Target variables and training data.}
Most of the vendors (15) offer custom or customizable assessments, adapting the
assessment to the client's particular data or job requirements. In practice, decisions about target variables and training data are made together based on where the data come from. Eight vendors build assessments based on data from the client's past and current employees (see
Figure~\ref{fig:pymetrics}). Vendors in general leave it up to clients to
determine what outcomes they want to predict, including, for example, performance reviews,
sales numbers, and retention time. Other vendors who offer customizable
assessments without using client data either use human expertise to determine
which of a pre-determined set of competencies are
most relevant to the particular job (the vendor's analysis of a job
role or a client's knowledge of relevant requirements) or don't explicitly specify their
prediction targets. In such cases, the vendor provides an assessment that scores applicants on various competencies, which are then combined
into a ``fit'' score based on a custom formula. Thus, even among vendors who
tailor their assessments to a client, they do so in different ways.

Vendors who only offer pre-built assessments typically either provide
assessments designed for a particular job role (e.g., salesperson), or provide a
sort of ``competency report'' with scores on a number of cognitive or
behavioral traits (e.g., leadership, grit, teamwork). These assessments are
closer in spirit to traditional psychometric assessments like the Myers-Briggs
Type Indicator or Big Five Personality Test; however, unlike traditional
assessments that rely on a small number of questions, modern
assessments may build psychographic profiles using
machine learning to analyze rich data sources like a video interview or gameplay.

\paragraph*{Validation.}
Generally, vendors' websites do not make clear whether vendors validate their models, what validation methodologies they use, how they select validation data, or how validation procedures might be tailored to the particular client.  Good \& Co.,\footnote{\url{https://good.co/}} notably, provides fairly
rigorous validation studies of the psychometric component of their assessment,
as well as a detailed audit of how the scores differ across demographic groups; however,
they do not provide similar documentation justifying the algorithmic techniques
they use to recommend candidates based on ``culture fit.''

\begin{table*}
  \centering
  \ra{1.4}
  \begin{tabularx}{.9\textwidth}{@{} l Y @{}}
    \toprule
    Vendor & Claim about bias \\
    \hline
    HireVue & Provide ``a highly valid, bias-mitigated assessment'' \\
    pymetrics & ``\ldots the Pre-Hire assessment does not show bias
    against women or minority respondents.'' \\
    PredictiveHire & ``AI bias is testable, hence fixable.'' \\
    Knockri & ``Knockri's A.I. is unbiased because of its full spectrum database that ensures there's no benchmark of what the `ideal candidate' looks like.'' \\
    \bottomrule
  \end{tabularx}
  \caption{Examples of claims that vendors make about bias, taken from
  their websites.}
  \label{tab:bias_claims}
\end{table*}

\paragraph{Accounting for bias.}
In total, while 15 of the vendors made at least abstract references to ``bias''
(sometimes in the context of well-established human bias in hiring), only 7
vendors explicitly discussed compliance or adverse impact with respect to the
assessments they offered. Three vendors explicitly mentioned the \nicefrac{4}{5}
rule, and an additional 4 advertised ``compliance'' or claimed to control
adverse impact more generally. Several of these vendors claimed to test models for
bias, ``fixing'' it when it appeared. HireVue and pymetrics, in particular, offered a detailed description of their overall approaches to de-biasing, which
involves removing features correlated with protected attributes when adverse
impact is detected. Other vendors (e.g., Knockri and PredictiveHire)
claimed to ``fix'' adverse impact when it is found without going into further detail.

Among those that do make concrete claims, all vendors we examined specifically
focus on equality of outcomes and compliance with the \nicefrac{4}{5} rule.
Roughly speaking, there are two ways in which vendors claim to achieve these
goals: naturally unbiased assessments and active algorithmic de-biasing.
Typically, vendors claiming to provide naturally unbiased assessments seek to
measure underlying cognitive or behavioral traits, so their assessments output a
small number of scores, one for each competency being measured. In this setting,
a naturally unbiased assessment in one that produces similar score distributions
across demographic groups. Koru, for instance, measures 7 traits (e.g., ``grit''
and ``presence'') and claims that ``[i]n all panels since 2015, the Pre-Hire
assessment does not show bias against women or minority
respondents''~\cite{jarrett2018koru}.

Other vendors actively intervene in their learned models to remove biases. One
technique that we have observed across multiple vendors (e.g., HireVue,
pymetrics, PredictiveHire) is the following: build a model and test it for
adverse impact against various subgroups.\footnote{pymetrics, for instance,
open-sources the tests it uses: \url{https://github.com/pymetrics/audit-ai}} As
Bogen and Rieke also observe~\cite{bogen2018help}, if adverse impact is found,
the model and/or data are modified to try to remove it, and then the model is
tested again for adverse impact.
HireVue and pymetrics downweight or remove features found to be highly
correlated with the protected attribute in question, noting that this can
significantly reduce adverse impact with little effect on the predictive
accuracy of the assessment.
This is done prior to the model's deployment on
actual applicants, though some vendors claim to periodically test and update
models.
In Section~\ref{sec:debiasing}, we discuss in depth
these efforts to define and remove bias.

\section{Analysis of Technical Concerns}
\label{sec:analysis}
Our findings in Section~\ref{sec:empirical} raise several technical challenges for the
pre-employment assessment process.
In this section, we focus on two areas that are particularly salient in the context of algorithmic hiring: \textbf{data choices}, where vendors must decide where to draw data from and what outcomes to predict; and the use of \textbf{alternative assessment formats}, like game- or video-based assessments that rely on larger feature sets and more complex machine learning tools than traditional question-based assessments.

\subsection{Data Choices}

Machine learning is often viewed as a process by which we predict a given output
from a given set of inputs. In reality, neither the inputs nor outputs are
fixed. Where do the data come from? What is the ``right'' outcome to predict?
These and others are crucial decisions in the machine learning pipeline, and can
create opportunities for bias to enter the process.

\paragraph*{Custom assessments.}
Consider a hypothetical practitioner who sets out to create a custom assessment
to determine who the ``best'' candidates are for her client. As is the case in
many domains, translating this to a feasible data-driven task forces our
practitioner to make certain compromises~\cite{passi2019problem}. It quickly
becomes clear that she must somehow operationalize ``best'' in some measurable
way. What does the client value? Sales numbers? Cultural fit? Retention? And,
crucially, what data does the client have? This is a nontrivial constraint: many
companies don't maintain comprehensive and accessible data about employee
performance, and thus, a practitioner may be forced to do the best she can with
the limited data that she is given~\cite{cappelli2018artificial}. Note that relying on the client's data has already
forced the practitioner to only learn from the client's existing employees; at
the outset, at least, she has no way to get data on how those who
\textit{weren't} hired would have performed.

Once a target is identified, the practitioner needs a dataset on which to train
a model. Since she has performance data on previous employees, she needs them to
take the assessment so she can link their assessment performance to their
observed job performance. How many employees' data does she need in order to get
an accurate model? What if certain employees don't want to or don't have time to
take the assessment? Is the set of employees who respond representative of the
larger applicant pool who will ultimately be judged based on this assessment?

Finally, the practitioner is in a position to actually build a model. Along the
way, however, she had to make several key choices, often based on factors (like
client data availability) outside her control. The choice of target variable is
particularly salient. Proxies like job evaluations, for instance,
can exhibit biases against
minorities~\cite{sidanius1989job,neumark1996sex,riach2002field}. Moreover,
predicting the success of future employees based on current employees
inherently skews the task toward finding candidates resembling those who have
already been hired.

Some vendors go beyond trying to identify candidates who are generically good,
or even good for a particular client, and explicitly focus on finding candidates
who ``fit'' with an existing employee or team. Both Good \& Co. and Teamscope
provide these team-specific tools for employers, and
Good \& Co. further advertises their assessments as a way to
``[r]eplicate your top performers.''\footnote{\url{https://good.co/pro/}} 
If models are localized to predict fit with particular teams or groups, any role
at any company could in principle have its own tailor-made predictive
model.
But when
models are fit and customized at such a small scale, it can be quite difficult
to determine what it means for such a model to be biased or discriminatory.
Does
each team-specific model need to be audited for bias? How would a vendor go about doing so?

And yet, while it is easy to criticize vendors for the choices they make, it's
not clear that there are better alternatives. In practice, it is impossible to
even define, let alone collect data on, an objective measure of a ``good''
employee. Nor is it always feasible to get data on a completely representative
sample of candidates. Vendors and advocates point out that many of the potentially
problematic elements here (subjective evaluations; biased historical samples;
emphasis on fit) are equally present, if not more so, in traditional human
hiring practices~\cite{chamorro2019should}.

\paragraph*{Customizable and pre-built assessments.}
Instead of building a new custom assessment for each client, it may be tempting
to instead offer a pre-built assessment (perhaps specific to a particular type
of job) that has been validated across data from a variety of clients. This has
the advantage that it isn't subject to the idiosyncratic data of each client; moreover, it can draw from a more diverse range of candidates and employees to learn a
broader notion of what a ``good'' employee looks like. Additionally, pre-built
assessments may be attractive to clients who do not have enough existing
employees from whom a custom assessment can be built.

\begin{figure*}[ht]
  \centering
  \includegraphics[width=.8\textwidth]{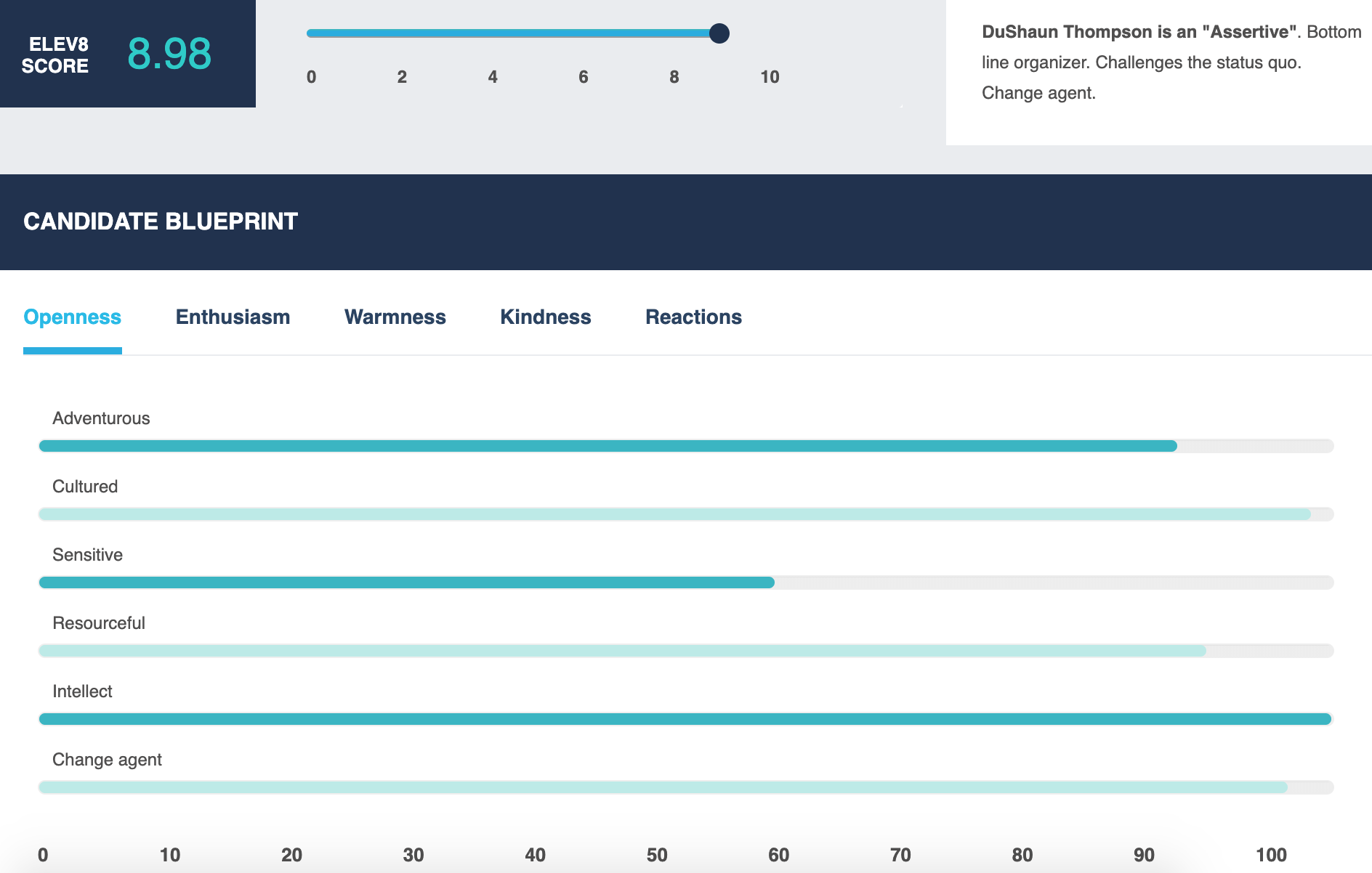}
  \caption{Part of a sample candidate profile from 8 and Above, based on a
    30-second recorded video cover letter
  (screenshot from the 8 and Above website: \url{https://www.8andabove.com/p/profile/blueprint/643})}
  \label{fig:8above}
\end{figure*}

Some vendors offer assessments that are mostly pre-built but somewhat
customizable. Koru and Plum.io, for example, provide pre-built assessments to
evaluate a fixed number of competencies. Experts then analyze the job
description and role for a particular client and determine which competencies
are most important for the client's needs. Thus, these vendors hope to get the
best of both worlds: assessments validated on large populations that are still
flexible enough to adapt to the specific requirements of each client. As shown
in Figure~\ref{fig:8above}, the firm 8 and Above profiles over 60 traits based on a video
interview, but also reports a single ``Elev8'' score tailored to the particular
client.

Despite these benefits, pre-built assessments do have drawbacks.
Individual competencies like ``grit'' or ``openness'' are themselves
constructs, and attempts to measure them must rely on other psychometric
assessments as ``ground truth.'' Given that traits can be measured by multiple
tests that don't perfectly correlate with one another~\cite{rodriguez2006meta},
it may be difficult to create an objective benchmark against which to compare an
algorithmic assessment. Furthermore, it is generally considered good practice
to build and validate assessments on a representative population for a
particular job role~\cite{society2003principles}, and both underlying candidate
pools and job specifics differ across locations, companies, and job
descriptions. Pre-built assessments must by nature be general, but as a
consequence, they may not adapt well to the client's requirements.

\paragraph*{Necessary trade-offs.}

This leads to an inherently challenging technical problem: on the one hand, more
data is usually beneficial in creating and validating an assessment; on the
other hand, drawing upon data from related but somewhat different sources may
lead to inaccurate conclusions. We can view this as an instance of domain adaptation and the
bias-variance tradeoff, well studied in the statistics and machine learning
literature~\cite{ben2010theory,friedman2001elements}. Pooling data from multiple companies
or geographic locations may reduce variance due to small sample sizes at a particular company, but comes at the cost of biasing the outcomes away from the client's specific needs.
There is no obvious answer or clear best practice here, and vendors and clients
must carefully consider the pros and cons of various assessment types. Larger
clients may be better positioned for vendors to build custom assessments based
solely on their data; smaller clients may turn to pre-built assessments, making
the assumption that the candidate pool and job role on which the assessment was
built is sufficiently similar to warrant generalizing its conclusions.

\subsection{Alternative Assessment Formats}
\label{sec:alt}
Once an assessment has been built, it must be validated to verify that it performs as expected.
Psychologists have developed extensive standards to guide assessment creators in this process~\cite{society2003principles}; however,
modern assessment vendors are pushing the boundaries of assessment formats far
beyond the pen-and-paper tests of old, often with little regulatory
oversight~\cite{chamorro2016new}. Game- and video-based assessments, in
particular, are increasingly common. Vendors point to an emerging line of literature showing that features derived from these modern assessment formats correlate with job outcomes and personality traits~\cite{kramer2010internal,grimmett2017veterinary} as evidence that these assessments contain information that can be predictive of job outcomes, though they rarely release rigorous validation studies of their own.

\paragraph*{Technical challenges for alternative assessments.}
While there is evidence for the predictive validity of alternative assessments,
empirical correlation is no substitute for theoretical justification.
Historically, IO psychologists have designed assessments based on their
research-driven knowledge that certain traits correlate with desirable outcomes.
To some extent, machine learning attempts to automate this process by
discovering relationships (e.g., between actions in a video game and personality
traits) instead of quantifying known relationships. Of course, machine learning
can be used to unearth meaningful relationships. But it may also find
relationships that experts don't understand. When the expert is unable to
explain why, for example, the cadence of a candidate's voice indicates higher
job performance, or why reaction time predicts employee retention, should a
vendor rely on these features? From a technical perspective, correlations that
cannot be theoretically justified may fail to generalize well or remain stable
over time, and, in light of such concerns, the APA Principles caution that a
practitioner should ``establish a clear rationale for linking the resulting
scores to the criterion constructs of interest''~\cite{society2003principles}.
Yet when an algorithm takes in ``millions of data points'' for each candidate
(as advertised by pymetrics\footnote{\url{https://perma.cc/3284-WTS8}}), it may
not be possible to provide a qualitative justification for the inclusion of each
feature.

Moreover, automated discovery of relationships makes it difficult for a critical expert to detect when the model makes indirect use of a proscribed characteristic. Rich sources of data can easily encode properties that are illegal to use in the hiring process.
Facial analysis, in particular, has been heavily scrutinized recently.
A wave of studies has shown that several commercially available facial
analysis techniques suffer from disparities in error rates across gender and
racial lines~\cite{buolamwini2018gender,raji2019actionable,rhue2018racial}, and
more broadly, evidence suggests that we may not be able to reliably infer emotions
from facial expressions, especially cross-culturally~\cite{barrett2019emotional}. Concerns have also been raised over the use of affect and emotion recognition for those with disabilities, particularly in the context of employment~\cite{fruchterman2018expanding,guo2019toward,hurley2016autism}.

Because it can be quite expensive and technically challenging to build facial
analysis software in-house, vendors will often turn to third parties (e.g.,
Affectiva\footnote{\url{https://www.affectiva.com/}}) who provide facial analysis as a
service. As a result, vendors lack the ability or resources to thoroughly audit
the software they use. With these concerns in mind, U.S. Senators Kamala Harris,
Patty Murray, and Elizabeth Warren recently wrote a letter to the EEOC asking
for a report on the legality and potential issues with the use of facial
analysis in pre-employment assessments~\cite{harris2018eeoc}.
Even more recently, Illinois passed a law requiring applicants to be notified
and provide consent if their video interviews will be analyzed by artificial
intelligence~\cite{ilbill}, though it's not clear what happens if an applicant
refuses to consent.

While heightened publicity regarding racial disparities in facial analysis has prompted many third-party vendors of this technology to respond by improving the performance of their tools on minority populations~\cite{puri2018mitigating,roach2018microsoft}, it remains unclear what information facial analysis relies on to draw conclusions about candidates.
Facial expressions may contain information about a range of sensitive attributes from obvious ones like ethnicity, gender, and age to more subtle traits like a candidate's mental and physical health~\cite{kramer2010internal,zhou2015tackling}.\footnote{As a general matter, the Americans with Disabilities Act prohibits employers from collecting or considering information about candidates' health~\cite{ada}.} Given the opacity of the deep learning models used for facial analysis, it can be difficult or even impossible to detect if a model inadvertently learns proxies for prohibited features.

\section{Algorithmic De-Biasing}
\label{sec:debiasing}
Under Title VII, employers bear ultimate legal responsibility for their hiring
decisions.
Employers, then, remain strongly motivated to mitigate their
potential liability against disparate impact claims. Vendors, in turn, are
incentivized to build demonstrably unbiased tools that help employers to avoid
such liability.

As we have described, all vendors in our sample who made concrete claims about de-biasing (including the two best-funded firms in our sample) did so with reference to equality of outcomes and compliance with the \nicefrac{4}{5} rule. In this section, we explore the effects of this reliance on the stages of a typical disparate impact lawsuit. We then explore technical approaches that have been proposed to control outcome disparities, and their relationship to the law. Finally, we describe some important consequences of the de-biasing strategies favored by vendors.

\subsection{Algorithmic De-Biasing and Disparate Impact Litigation} 

Recall the three steps in a disparate impact case. The plaintiff must first establish that the employer's selection procedure generates a disparate impact. Once established, the employer must then defend itself by justifying the disparate impact by reference to some business necessity. In this case, an employer would likely do so by establishing the validity of the model driving its hiring decisions. Finally, the plaintiff may then challenge the proffered justification as faulty or demonstrate that an alternative practice exists that would serve the employer's business objective equally well while reducing the disparate impact in its selection rates.

Note that disparate impact doctrine does not prohibit disparate impact altogether; it renders employers liable for an \textit{unjustified} or \textit{avoidable} disparate impact. Vendors' choice to enforce the \nicefrac{4}{5} rule might therefore seem overly cautious: although employers could justify an assessment that has a disparate impact by demonstrating its validity (as we discuss in Section 2), vendors take steps to ensure that employers are not placed in this position, because assessments are prevented from having a disparate impact in the first place. One possible explanation for adopting the \nicefrac{4}{5} rule is that vendors might be catering to employers' aversion to legal risk.

As to the second step, the practical effect of vendors' reliance on the
\nicefrac{4}{5} rule is to obviate the need for an employer to demonstrate
business necessity through a legally rigorous validation process. According to
the Uniform Guidelines, employers only need to validate their selection
procedure if it has a disparate impact. Of course, clients might still expect
and even demand validation studies from vendors, given their goal of selecting
qualified candidates. As a consequence, the choice of how to validate seems to
become a \textit{business} decision rather than a \textit{legal} imperative. 

The final step in a disparate impact case raises yet another possible explanation for vendors' decisions to adopt the \nicefrac{4}{5} rule as a constraint. Recall that, at this stage, employers bear liability if they failed to adopt an alternative practice that could have minimized any business-justified disparity created by their selection procedure, provided that such practices were not too costly. Employers therefore run significant legal risks if they do not take such steps.  In turn, should vendors have some way to minimize disparity without sacrificing the accuracy of their assessments, failing to do so might place their clients in legal jeopardy. A plaintiff could assert that this very possibility reveals that any evident disparate impact---even if justified by a validation study---was avoidable.

While the burden of identifying this alternative business practice rests with
the plaintiff, vendors may want to preempt this argument by taking affirmative
steps to explore how to minimize disparate impact without imposing unwelcome
costs on the employer. In the past, such exploratory efforts might have been
costly and difficult, since discovering an alternative business practice that is
equally effective for the firm, while generating less disparity in selection
rates, was no easy task.
Many modern assessments (e.g., those with a large number of features)
make some degree of exploration almost trivial, allowing vendors to find a model
that (nearly) maintains maximum accuracy while reducing disparate impact.

In this way, the ready availability of algorithmic techniques might effectively
create a legal \textit{imperative} to use them. If the adverse impact of a
business-justified model could be reduced through algorithmic
de-biasing---without significantly harming predictive ability, and at trivial
cost---de-biasing itself might be considered an ``alternative business
practice,'' and therefore render the employer liable for not adopting it. 

\subsection{Methods to Control Outcome Disparities}
Thus, for legal reasons, a vendor may choose to control outcome disparities in strict adherence to the \nicefrac{4}{5} rule.
But this is not the end of the story; multiple techniques exist to control outcome differences.
Here, we explore both historical and contemporary approaches in comparison with the de-biasing techniques we observe.

The most straightforward approach to control outcome differences is known as
``within-group scoring,'' under which scores are reported as a percentile with
respect to the particular group in question. Employers could then select
candidates above a particular threshold for each group (top 10\% from Group A,
top 10\% from Group B, etc.), which would naturally result in equal selection
rates. Recall that in the de-biasing reviewed above, vendors achieve
(approximately) equal selection rates by systematically removing features from
the model that contribute to a disparate impact. In so doing, they may lose
useful information contained in these features as well, undermining their
ability to maintain an accurate rank order within each group. In contrast,
within-group scoring may theoretically be the optimal way to equalize selection
rates, since it preserves rank
order~\cite{corbett2017algorithmic,lipton2018does}.

In fact, within-group scoring was used for the General Aptitude Test Battery
(GATB), a pre-employment assessment developed in the 1940s by the US Employment
Service (USES), due to significant differences in score distributions across
ethnic groups. In particular, the USES reported results as within-group
percentile scores by ethnicity---black, Hispanic, and
other~\cite{national1989fairness,schuler2013personnel}. Commissioned to
investigate the justification for such a policy, a National Academy of Sciences
study recommended the continued use of within-group percentiles because without
them, minority applicants would suffer from ``higher false-rejection
rates''~\cite{national1989fairness}.

In principle, within-group score reporting (also known as ``race-norming'')
would satisfy the \nicefrac{4}{5} rule; so why don't vendors use it? In fact,
within-group reporting would likely be considered illegal today. In 1986 the
Department of Justice challenged its legality in the GATB, claiming that it
constituted disparate treatment~\cite{schuler2013personnel}, and the practice
was prohibited by the Civil Rights Act of 1991~\cite{civil1991}.

This points to a longstanding tension between disparate treatment and disparate impact: some techniques to control outcome disparities require the use of protected attributes, which may be considered disparate treatment.
To circumvent this, the vendors we observe engaging in algorithmic de-biasing take into account protected attributes when \textit{building} models, but ultimately produce models that do not take protected attributes as input.
In this way, individual decisions do not exhibit disparate treatment, and yet, outcome disparities can still be mitigated.

In fact, these techniques fit into a broader category of methods known as
Disparate Learning Processes (DLPs), a family of algorithms designed to produce
decision rules that (approximately) equalize outcomes without engaging in
disparate treatment at the individual
level~\cite{lipton2018does,pedreshi2008discrimination,zafar2015fairness}. There
are slight differences between DLPs as found in the computer science literature
and vendors' algorithmic de-biasing efforts: DLPs typically work by imposing
constraints that prevent outcome disparities on the learning algorithm that
produces the model; the algorithmic de-biasing we observe, on the other hand,
simply removes features correlated with protected attributes until outcomes are
within a tolerable range. In spirit, however, these techniques are ultimately
quite related.

Similar connections exist to ``fair representation'' learning, where an ``encoder'' is built to process data by removing information about protected attributes, including proxies and correlations~\cite{zemel2013learning,madras2018learning,edwards2015censoring}.
Thus, any model built on data processed by the encoder would have approximately equal outcomes, since outputs of the encoder contain very little information about protected attributes.
As in DLPs, protected attributes are used only to create the encoder; after deployment, when the encoder processes any individual's data, it does not have access to protected attributes.
We can think of some vendors' practices as analogous to building such an encoder---one that ``processes'' data by simply discarding features highly correlated with protected attributes.

\subsection{Limitations of Outcome-Based De-Biasing}
Despite the perhaps good reasons vendors have to use the particular form of
algorithmic de-biasing discussed above, these techniques face important caveats
and consequences worth mentioning.

Outcome-based notions of bias are
intimately tied to the datasets on which they are evaluated. As both the EEOC
Guidelines and APA Principles clearly articulate, a representative sample is
crucial for validation~\cite{equal1978department,society2003principles}. The
same holds true for claims regarding outcome disparities: they may depend on whether the assessment is taken by recent college
grads in Michigan applying for sales positions or high school dropouts
in New York applying for jobs stocking warehouses.
Thus, when evaluating claims regarding outcome disparities, it is critical to understand how vendors collect and maintain relevant, representative data.

While outcome disparities are important for vendors to consider, especially in
light of U.S. regulations, discrimination and the \nicefrac{4}{5} rule should
not be conflated. Vendors may find it necessary from a legal or business
perspective to build models that satisfy the \nicefrac{4}{5} rule, but this is
not a substitute for a critical analysis into the mechanisms by which bias and
harm manifest in an assessment. For example, differential validity, which occurs
when an assessment is better at ranking members of one group than another,
should be a top-level concern when examining an
assessment~\cite{society2003principles,young2001differential}. But because of
the legal emphasis placed on adverse impact, vendors have little incentive to
structure their techniques around it. Furthermore, it can be challenging to
identify and mitigate outcome disparities with respect to protected attributes
employers typically don't collect (e.g., sexual orientation~\cite{feha}).
In such cases, vendors may need to consider alternative approaches to
prevent discrimination.

More broadly, bias is not limited to the task of predicting outputs from inputs;
a critical, holistic examination of the entire assessment development pipeline may surface deeper concerns. 
Where do inputs and outputs come from, and what justification do they
have? Are there features that shouldn't be used? This isn't to say that some
vendors are not already asking these questions; however, in the interest of
forming industry standards surrounding algorithmic assessments,
the legal operationalization of the \nicefrac{4}{5} rule as a definition of bias runs the risk of
downplaying the importance of examining a system as a whole.

Both the law and existing techniques focus on assessment outcomes as binary
(screened in/out); however, some platforms actually rank candidates (explicitly,
or implicitly by assigning numerical scores). While screening decisions can
ultimately be viewed as binary (a candidate is either interviewed or not), there
are a number of subtleties induced by ranking: a lower-ranked candidate may only
be interviewed after higher-ranked candidates, and their lower score could
unduly bias future decision-makers against them~\cite{bogen2018help}. There is
no clear analog of the \nicefrac{4}{5} rule for ranking; in practice, vendors
may choose a cut-off score and test for adverse impact via the \nicefrac{4}{5}
rule~\cite{equal1978department,auditai}. In the computer science literature,
there are ongoing efforts to define technical constraints on rankings in the
spirit of equal representation and the \nicefrac{4}{5}
rule~\cite{celis2017ranking,yang2017measuring,zehlike2017fa}, and LinkedIn has
adopted a similar approach to encourage demographic diversity in its search
results~\cite{geyik2019fairness}. However, none of these approaches has received
any sort of consensus or official endorsement.

From a policy perspective, the EEOC can and should clarify its position on the
use of algorithmic de-biasing techniques, which to our knowledge has yet to be challenged in court.
Legal scholars have begun to debate the legality of ``algorithmic affirmative action'' in various contexts~\cite{kroll2016accountable,bent2019algorithmic,hellman2019measuring,kim2017auditing,raub2018bots}, but the debate is far from settled.
While existing guidelines can be argued to
apply to ML-based assessments, the de-biasing techniques described above do
present new opportunities and challenges.

\section{Discussion and Recommendations}
\label{sec:discussion}
In this work, we have presented an in-depth analysis into the bias-related practices of vendors of algorithmic pre-employment assessments.
Our findings have implications not only for hiring pipelines, but more broadly for investigations into algorithmic and socio-technical systems.
Given the proprietary and sensitive nature of models built for actual clients, it is often infeasible for external researchers to perform a traditional audit; despite this, we are able to glean valuable information by delving into vendors' publicly available statements.
Broadly speaking, models result from the application of a \textbf{vendor's practices} to a real-world setting.
Thus, by learning about these practices, we can draw conclusions and raise relevant questions about the resultant models.
In doing so, we can create a common vocabulary with which we can discuss and compare practices.
We found it useful to \textbf{limit the scope} of our inquiry in order to be able to ask and answer concrete questions.
Even just considering algorithms used in the context of hiring, we found enough heterogeneity (as have previous reports on the subject~\cite{bogen2018help,garr2019diversity}) that it was necessary to further refine our focus to those used in pre-employment assessments.
While this did lead us to exclude a number of innovative and intriguing hiring
technologies (see, e.g., Textio\footnote{Textio (\url{https://textio.com/})
analyzes job descriptions for gender bias and makes suggestions for alternative,
gender-neutral framings.} or Jopwell\footnote{Jopwell
(\url{https://www.jopwell.com/}) builds and maintains a network of Black,
Latinx, and Native American students and connects students these with
employers.}), it allowed us to make specific and direct comparisons between
vendors and get a more detailed understanding of the technical challenges
specific to assessments.

In analyzing models via practices, we observe that it is crucial to consider
technical systems in conjunction with the \textbf{context} surrounding their use
and deployment. It would be difficult to understand vendors' design decisions
without paying attention to the relevant legal, historical, and social
influences. Moreover, in order to push beyond hypothetical or anecdotal accounts
of algorithmic bias, we need to incorporate empirical evidence from the field.

Based on our findings, we summarize the following policy recommendations
discussed throughout this work.
\begin{enumerate}
  \item Transparency is crucial to further our understanding of these systems.
    While there are some exceptions, vendors in general are not particularly
    forthcoming about their practices. Additional transparency is necessary to
    craft effective policy and enable meaningful oversight.
  \item Disparate impact is not the only indicator of bias. Vendors should also
    monitor other metrics like differential validity.
  \item Outcome-based measures of bias (including tests for disparate
    impact and differential validity) are limited in their power. They require
    representative datasets for particular applicant pools and fail to
    critically examine the appropriateness of individual predictors. Moreover,
    they depend on access to protected attributes that are not always available.
  \item We may need to reconsider legal standards of validity under the Uniform
    Guidelines in light of machine learning. Because machine learning may
    discover relationships that we do not understand, a statistically valid
    assessment may inadvertently leverage ethically problematic correlations.
  \item Algorithmic de-biasing techniques have significant implications for
    ``alternative business practices,'' since they automate the search for less
    discriminatory alternatives. Vendors should explore these techniques to
    reduce disparate impact, and the EEOC should offer clarity about how the law applies.
\end{enumerate}

Our work leads naturally to a range of questions, ranging from those that seem
quite technical (What is the effect of algorithmic de-biasing on model outputs?
When should data from other sources be incorporated?) to socio-political (What
additional regulatory constraints could improve the use of algorithms in
assessment? How can assessments promote the autonomy and dignity of
candidates?).
Because the systems we examine are shaped by technical, legal, political, and
social forces, we believe that an interdisciplinary approach is necessary to get
a broader picture of both the problems they face and the potential avenues for
improvement.

\paragraph*{Acknowledgments}
We thank Rediet Abebe, Ifeoma Ajunwa, Bilan Ali, Lewis Baker, Miranda Bogen,
Heather Bussing, Albert Chang, A. F. Cooper, Fernando Delgado, Kate Donahue,
Stacia Garr, Avital Gertner-Samet, Jim Guszcza, Stephen Hilgartner, Lauren
Kilgour, Loren Larsen, Richard Marr, Cassidy McGovern, Helen Nissenbaum, Samir
Passi, David Pedulla, Frida Polli, Sarah Riley, David Robinson, Caleb Rottman,
John Sumser, Kelly Trindel, Katie Van Koevering, Briana Vecchione, Suresh
Venkatasubramanian, Angela Zhou, Malte Ziewitz, and Lindsey Zuloaga for their
suggestions and insights.

\bibliographystyle{plain}
\bibliography{refs}

\begin{thebibliography}{10}

\bibitem{ajunwa2019paradox}
Ifeoma Ajunwa.
\newblock The paradox of automation as anti-bias intervention.
\newblock {\em Cardozo Law Review}, 41, 2020.

\bibitem{angwin2016machine}
Julia Angwin, Jeff Larson, Surya Mattu, and Lauren Kirchner.
\newblock Machine bias.
\newblock {\em ProPublica, May}, 23, 2016.

\bibitem{ilbill}
Illinois~General Assembly.
\newblock Artificial intelligence video interview act, 2019.

\bibitem{auditai}
Lewis Baker, David Weisberger, Daniel Diamond, Mark Ward, and Joe Naso.
\newblock audit-{AI}.
\newblock \url{https://github.com/pymetrics/audit-ai}, 2018.

\bibitem{baritz1960servants}
Loren Baritz.
\newblock {\em The servants of power: A history of the use of social science in
  American industry.}
\newblock Wesleyan University Press, 1960.

\bibitem{barocas2016big}
Solon Barocas and Andrew~D Selbst.
\newblock Big data's disparate impact.
\newblock {\em Calif. L. Rev.}, 104:671, 2016.

\bibitem{barrett2019emotional}
Lisa~Feldman Barrett, Ralph Adolphs, Stacy Marsella, Aleix~M Martinez, and
  Seth~D Pollak.
\newblock Emotional expressions reconsidered: Challenges to inferring emotion
  from human facial movements.
\newblock {\em Psychological science in the public interest}, 20(1):1--68,
  2019.

\bibitem{ben2010theory}
Shai Ben-David, John Blitzer, Koby Crammer, Alex Kulesza, Fernando Pereira, and
  Jennifer~Wortman Vaughan.
\newblock A theory of learning from different domains.
\newblock {\em Machine learning}, 79(1-2):151--175, 2010.

\bibitem{bendick2012developing}
Marc Bendick and Ana~P Nunes.
\newblock Developing the research basis for controlling bias in hiring.
\newblock {\em Journal of Social Issues}, 68(2):238--262, 2012.

\bibitem{bendick1997employment}
Marc Bendick~Jr, Charles~W Jackson, and J~Horacio Romero.
\newblock Employment discrimination against older workers: An experimental
  study of hiring practices.
\newblock {\em Journal of Aging \& Social Policy}, 8(4):25--46, 1997.

\bibitem{bent2019algorithmic}
Jason~R Bent.
\newblock Is algorithmic affirmative action legal?
\newblock {\em Georgetown Law Journal}, 108, 2020.

\bibitem{bertrand2004emily}
Marianne Bertrand and Sendhil Mullainathan.
\newblock Are {E}mily and {G}reg more employable than {L}akisha and {J}amal?
  {A} field experiment on labor market discrimination.
\newblock {\em American economic review}, 94(4):991--1013, 2004.

\bibitem{biddle2008uniform}
Daniel~A Biddle.
\newblock Are the uniform guidelines outdated? federal guidelines, professional
  standards, and validity generalization (vg).
\newblock {\em The Industrial-Organizational Psychologist}, 45(4):17--23, 2008.

\bibitem{bogen2018help}
Miranda Bogen and Aaron Rieke.
\newblock Help wanted: An exploration of hiring algorithms, equity, and bias.
\newblock Technical report, Upturn, 2018.

\bibitem{bornstein2018antidiscriminatory}
Stephanie Bornstein.
\newblock Antidiscriminatory algorithms.
\newblock {\em Ala. L. Rev.}, 70:519, 2018.

\bibitem{buolamwini2018gender}
Joy Buolamwini and Timnit Gebru.
\newblock Gender shades: Intersectional accuracy disparities in commercial
  gender classification.
\newblock In {\em Conference on Fairness, Accountability and Transparency},
  pages 77--91, 2018.

\bibitem{cappelli2018artificial}
Peter Cappelli, Prasanna Tambe, and Valery Yakubovich.
\newblock Artificial intelligence in human resources management: Challenges and
  a path forward.
\newblock {\em Available at SSRN 3263878}, 2018.

\bibitem{celis2017ranking}
L.~Elisa Celis, Damian Straszak, and Nisheeth~K. Vishnoi.
\newblock Ranking with fairness constraints.
\newblock In {\em 45th International Colloquium on Automata, Languages, and
  Programming, {ICALP} 2018, July 9-13, 2018, Prague, Czech Republic}, pages
  28:1--28:15, 2018.

\bibitem{chamorro2016new}
Tomas Chamorro-Premuzic, Dave Winsborough, Ryne~A Sherman, and Robert Hogan.
\newblock New talent signals: Shiny new objects or a brave new world?
\newblock {\em Industrial and Organizational Psychology}, 9(3):621--640, 2016.

\bibitem{chamorro2019should}
Tomas Chamorro-Prezumic and Reece Akhtar.
\newblock Should companies use {AI} to assess job candidates?
\newblock {\em Harvard Business Review}, 2019.

\bibitem{cohn1979use}
Richard~M Cohn.
\newblock On the use of statistics in employment discrimination cases.
\newblock {\em Ind. LJ}, 55:493, 1979.

\bibitem{cohn1979statistical}
Richard~M Cohn.
\newblock Statistical laws and the use of statistics in law: A rejoinder to
  {Professor Shoben}.
\newblock {\em Ind. LJ}, 55:537, 1979.

\bibitem{equal1978department}
Equal Employment~Opportunity Commission, Civil~Service Commission, et~al.
\newblock Uniform guidelines on employee selection procedures.
\newblock {\em Federal Register}, 43(166):38290--38315, 1978.

\bibitem{titlevii}
U.S. Congress.
\newblock Civil rights act, 1964.

\bibitem{ada}
U.S. Congress.
\newblock Americans with disabilities act, 1990.

\bibitem{civil1991}
U.S. Congress.
\newblock Civil rights act, 1991.

\bibitem{corbett2017algorithmic}
Sam Corbett-Davies, Emma Pierson, Avi Feller, Sharad Goel, and Aziz Huq.
\newblock Algorithmic decision making and the cost of fairness.
\newblock In {\em Proceedings of the 23rd ACM SIGKDD International Conference
  on Knowledge Discovery and Data Mining}, pages 797--806. ACM, 2017.

\bibitem{national1989fairness}
National~Research Council et~al.
\newblock {\em Fairness in employment testing: Validity generalization,
  minority issues, and the General Aptitude Test Battery}.
\newblock National Academies Press, 1989.

\bibitem{national2013new}
National~Research Council et~al.
\newblock {\em New directions in assessing performance potential of individuals
  and groups: Workshop summary}.
\newblock National Academies Press, 2013.

\bibitem{cowgill2018bias}
Bo~Cowgill.
\newblock Bias and productivity in humans and algorithms: Theory and evidence
  from resume screening.
\newblock {\em Columbia Business School, Columbia University}, 29, 2018.

\bibitem{cravens1978triumph}
Hamilton Cravens.
\newblock {\em The triumph of evolution: The heredity--environment controversy,
  1900--1941.}
\newblock Johns Hopkins University Press, 1978.

\bibitem{dubois1970history}
Philip~Hunter DuBois.
\newblock {\em A history of psychological testing}.
\newblock Allyn and Bacon, 1970.

\bibitem{dunnette1979personnel}
Marvin~D Dunnette and Walter~C Borman.
\newblock Personnel selection and classification systems.
\newblock {\em Annual review of psychology}, 30(1):477--525, 1979.

\bibitem{edwards2015censoring}
Harrison Edwards and Amos~J. Storkey.
\newblock Censoring representations with an adversary.
\newblock In {\em 4th International Conference on Learning Representations,
  {ICLR} 2016, San Juan, Puerto Rico, May 2-4, 2016, Conference Track
  Proceedings}, 2016.

\bibitem{feldman2015certifying}
Michael Feldman, Sorelle~A Friedler, John Moeller, Carlos Scheidegger, and
  Suresh Venkatasubramanian.
\newblock Certifying and removing disparate impact.
\newblock In {\em Proceedings of the 21th ACM SIGKDD International Conference
  on Knowledge Discovery and Data Mining}, pages 259--268. ACM, 2015.

\bibitem{society2003principles}
Society for Industrial, Organizational~Psychology (US), and American
  Psychological Association.~Division of~Industrial-Organizational~Psychology.
\newblock {\em Principles for the validation and use of personnel selection
  procedures}.
\newblock American Psychological Association, 2018.

\bibitem{friedman2001elements}
Jerome Friedman, Trevor Hastie, and Robert Tibshirani.
\newblock {\em The elements of statistical learning}.
\newblock Springer series in statistics New York, 2001.

\bibitem{fruchterman2018expanding}
Jim Fruchterman and Joan Melllea.
\newblock Expanding employment success for people with disabilities.
\newblock Technical report, benetech, 2018.

\bibitem{garr2019diversity}
Stacia~Sherman Garr and Carole Jackson.
\newblock Diversity \& inclusion technology: The rise of a transformative
  market.
\newblock Technical report, RedThread Research, 2019.

\bibitem{gerhardt1916scientific}
PW~Gerhardt.
\newblock Scientific selection of employees.
\newblock {\em Electric Railway Journal}, 47, 1916.

\bibitem{geyik2019fairness}
Sahin~Cem Geyik, Stuart Ambler, and Krishnaram Kenthapadi.
\newblock Fairness-aware ranking in search \& recommendation systems with
  application to linkedin talent search.
\newblock In {\em Proceedings of the 25th ACM SIGKDD International Conference
  on Knowledge Discovery and Data Mining}. ACM, 2019.

\bibitem{grimmett2017veterinary}
Jeff Grimmett.
\newblock Veterinary practitioners - personal characteristics and professional
  longevity.
\newblock {\em VetScript}, 2017.

\bibitem{guo2019toward}
Anhong Guo, Ece Kamar, Jennifer~Wortman Vaughan, Hannah Wallach, and
  Meredith~Ringel Morris.
\newblock Toward fairness in ai for people with disabilities: A research
  roadmap.
\newblock {\em ACM SIGACCESS}, 125, October 2019.

\bibitem{guzzo2015big}
Richard~A Guzzo, Alexis~A Fink, Eden King, Scott Tonidandel, and Ronald~S
  Landis.
\newblock Big data recommendations for industrial--organizational psychology.
\newblock {\em Industrial and Organizational Psychology}, 8(4):491--508, 2015.

\bibitem{haney1982employment}
Craig Haney.
\newblock Employment tests and employment discrimination: A dissenting
  psychological opinion.
\newblock {\em Indus. Rel. LJ}, 5:1, 1982.

\bibitem{harris2018eeoc}
Kamala~D. Harris, Patty Murray, and Elizabeth Warren.
\newblock Letter to {U}.{S}. {E}qual {E}mployment {O}pportunity {C}ommission,
  2018.

\bibitem{hellman2019measuring}
Deborah Hellman.
\newblock Measuring algorithmic fairness.
\newblock {\em Virginia Public Law and Legal Theory Research Paper}, (2019-39),
  2019.

\bibitem{houser2019can}
Kimberly Houser.
\newblock Can {AI} solve the diversity problem in the tech industry? mitigating
  noise and bias in employment decision-making.
\newblock {\em Stanford Technology Law Review}, 22, 2019.

\bibitem{hurley2016autism}
Amy~E. Hurley-Hanson and Cristina~M. Giannantonio.
\newblock Autism in the workplace.
\newblock In {\em Journal of Business Management}, volume~22, 2016.

\bibitem{hutchinson201950}
Ben Hutchinson and Margaret Mitchell.
\newblock 50 years of test (un) fairness: Lessons for machine learning.
\newblock In {\em Proceedings of the Conference on Fairness, Accountability,
  and Transparency}, pages 49--58. ACM, 2019.

\bibitem{jarrett2018koru}
Josh Jarrett and Sarah Croft.
\newblock The science behind the {K}oru model of predictive hiring for fit.
\newblock Technical report, Koru, 2018.

\bibitem{johnson2016if}
Stefanie~K Johnson, David~R Hekman, and Elsa~T Chan.
\newblock If there's only one woman in your candidate pool, there's
  statistically no chance she'll be hired.
\newblock {\em Harvard Business Review}, 26(04), 2016.

\bibitem{kemble1916testing}
William~F Kemble.
\newblock Testing the fitness of your employees.
\newblock {\em Industrial Management}, 1916.

\bibitem{kim2016data}
Pauline~T Kim.
\newblock Data-driven discrimination at work.
\newblock {\em Wm. \& Mary L. Rev.}, 58:857, 2016.

\bibitem{kim2017auditing}
Pauline~T Kim.
\newblock Auditing algorithms for discrimination.
\newblock {\em U. Pa. L. Rev. Online}, 166:189, 2017.

\bibitem{kim2018big}
Pauline~T Kim.
\newblock Big data and artificial intelligence: New challenges for workplace
  equality.
\newblock {\em U. Louisville L. Rev.}, 57:313, 2018.

\bibitem{kim2020manipulating}
Pauline~T Kim.
\newblock Manipulating opportunity.
\newblock {\em Virginia Law Review}, 106, 2020.

\bibitem{kleinberg2017human}
Jon Kleinberg, Himabindu Lakkaraju, Jure Leskovec, Jens Ludwig, and Sendhil
  Mullainathan.
\newblock Human decisions and machine predictions.
\newblock {\em The Quarterly Journal of Economics}, 133(1):237--293, 2017.

\bibitem{kleinberg2019discrimination}
Jon Kleinberg, Jens Ludwig, Sendhil Mullainathan, and Cass~R Sunstein.
\newblock Discrimination in the age of algorithms.
\newblock {\em Journal of Legal Analysis}, 2019.

\bibitem{kramer2010internal}
Robin~SS Kramer and Robert Ward.
\newblock Internal facial features are signals of personality and health.
\newblock {\em The Quarterly Journal of Experimental Psychology},
  63(11):2273--2287, 2010.

\bibitem{kroll2016accountable}
Joshua~A Kroll, Solon Barocas, Edward~W Felten, Joel~R Reidenberg, David~G
  Robinson, and Harlan Yu.
\newblock Accountable algorithms.
\newblock {\em U. Pa. L. Rev.}, 165:633, 2016.

\bibitem{feha}
California~State Legislature.
\newblock Fair employment and housing act, 1959.

\bibitem{lipton2018does}
Zachary Lipton, Julian McAuley, and Alexandra Chouldechova.
\newblock Does mitigating ml's impact disparity require treatment disparity?
\newblock In {\em Advances in Neural Information Processing Systems}, pages
  8125--8135, 2018.

\bibitem{madaus2001adverse}
George~F Madaus and Marguerite Clarke.
\newblock The adverse impact of high stakes testing on minority students:
  Evidence from 100 years of test data.
\newblock Technical report, ERIC, 2001.

\bibitem{madras2018learning}
David Madras, Elliot Creager, Toniann Pitassi, and Richard Zemel.
\newblock Learning adversarially fair and transferable representations.
\newblock In {\em Proceedings of the 35th International Conference on Machine
  Learning}, volume~80, pages 3384--3393, Stockholmsmässan, Stockholm Sweden,
  10--15 Jul 2018. PMLR.

\bibitem{mariotti2017talent}
Andrew Mariotti.
\newblock Talent acquisition benchmarking report.
\newblock Technical report, Society for Human Resource Management, 2017.

\bibitem{mcdaniel2011uniform}
Michael~A Mcdaniel, Sven Kepes, and George~C Banks.
\newblock The uniform guidelines are a detriment to the field of personnel
  selection.
\newblock {\em Industrial and Organizational Psychology}, 4(4):494--514, 2011.

\bibitem{munsterberg1998psychology}
Hugo Munsterberg.
\newblock {\em Psychology and industrial efficiency}, volume~49.
\newblock A\&C Black, 1998.

\bibitem{myers1962myers}
Isabel~Briggs Myers.
\newblock {\em The {M}yers-{B}riggs type indicator}.
\newblock Consulting Psychologists Press, 1962.

\bibitem{neumark1996sex}
David Neumark, Roy~J Bank, and Kyle~D Van~Nort.
\newblock Sex discrimination in restaurant hiring: An audit study.
\newblock {\em The Quarterly journal of economics}, 111(3):915--941, 1996.

\bibitem{norman1963toward}
Warren~T Norman.
\newblock Toward an adequate taxonomy of personality attributes: Replicated
  factor structure in peer nomination personality ratings.
\newblock {\em The Journal of Abnormal and Social Psychology}, 66(6):574, 1963.

\bibitem{passi2019problem}
Samir Passi and Solon Barocas.
\newblock Problem formulation and fairness.
\newblock In {\em Proceedings of the Conference on Fairness, Accountability,
  and Transparency}, pages 39--48. ACM, 2019.

\bibitem{pedreshi2008discrimination}
Dino Pedreshi, Salvatore Ruggieri, and Franco Turini.
\newblock Discrimination-aware data mining.
\newblock In {\em Proceedings of the 14th ACM SIGKDD international conference
  on Knowledge discovery and data mining}, pages 560--568. ACM, 2008.

\bibitem{puri2018mitigating}
Ruchir Puri.
\newblock Mitigating bias in {AI} models.
\newblock {\em IBM Research Blog}, 2018.

\bibitem{quillian2017meta}
Lincoln Quillian, Devah Pager, Ole Hexel, and Arnfinn~H Midtb{\o}en.
\newblock Meta-analysis of field experiments shows no change in racial
  discrimination in hiring over time.
\newblock {\em Proceedings of the National Academy of Sciences},
  114(41):10870--10875, 2017.

\bibitem{raji2019actionable}
Inioluwa~Deborah Raji and Joy Buolamwini.
\newblock Actionable auditing: Investigating the impact of publicly naming
  biased performance results of commercial {AI} products.
\newblock {\em AAAI/ACM Conf. on AI Ethics and Society}, 2019.

\bibitem{raub2018bots}
McKenzie Raub.
\newblock Bots, bias and big data: Artificial intelligence, algorithmic bias
  and disparate impact liability in hiring practices.
\newblock {\em Ark. L. Rev.}, 71:529, 2018.

\bibitem{rhue2018racial}
Lauren Rhue.
\newblock Racial influence on automated perceptions of emotions.
\newblock {\em Available at SSRN 3281765}, 2018.

\bibitem{riach2002field}
Peter~A Riach and Judith Rich.
\newblock Field experiments of discrimination in the market place.
\newblock {\em The economic journal}, 112(483):F480--F518, 2002.

\bibitem{roach2018microsoft}
John Roach.
\newblock Microsoft improves facial recognition technology to perform well
  across all skin tones, genders.
\newblock {\em The AI Blog}, 2018.

\bibitem{rodriguez2006meta}
Michael~C Rodriguez and Yukiko Maeda.
\newblock Meta-analysis of coefficient alpha.
\newblock {\em Psychological methods}, 11(3):306, 2006.

\bibitem{ruda1968racial}
Edward Ruda and Lewis~E Albright.
\newblock Racial differences on selection instruments related to subsequent job
  performance.
\newblock {\em Personnel Psychology}, 1968.

\bibitem{salas2011}
Eduardo Salas.
\newblock Reply to request for public comment on plan for retrospective
  analysis of significant regulations pursuant to executive order 13563, 2011.

\bibitem{sanchez2019does}
Javier Sanchez-Monedero, Lina Dencik, and Lilian Edwards.
\newblock What does it mean to solve the problem of discrimination in hiring?
  social, technical and legal perspectives from the uk on automated hiring
  systems.
\newblock In {\em Proceedings of the Conference on Fairness, Accountability,
  and Transparency}. ACM, 2020.

\bibitem{schuler2013personnel}
Heinz Schuler, James~L Farr, and Mike Smith.
\newblock {\em Personnel selection and assessment: Individual and
  organizational perspectives}.
\newblock Psychology Press, 1993.

\bibitem{shoben1978differential}
Elaine~W Shoben.
\newblock Differential pass-fail rates in employment testing: Statistical proof
  under {Title VII}.
\newblock {\em Harvard Law Review}, pages 793--813, 1978.

\bibitem{shoben1979defense}
Elaine~W Shoben.
\newblock In defense of disparate impact analysis under {Title VII}: A reply to
  {D}r. {C}ohn.
\newblock {\em Ind. LJ}, 55:515, 1979.

\bibitem{sidanius1989job}
Jim Sidanius and Marie Crane.
\newblock Job evaluation and gender: The case of university faculty.
\newblock {\em Journal of Applied Social Psychology}, 19(2):174--197, 1989.

\bibitem{terman1916measurement}
Lewis~Madison Terman.
\newblock {\em The measurement of intelligence: An explanation of and a
  complete guide for the use of the Stanford revision and extension of the
  Binet-Simon intelligence scale}.
\newblock Houghton Mifflin, 1916.

\bibitem{tyler1947psychology}
Leona~E Tyler.
\newblock {\em The psychology of human differences.}
\newblock D Appleton-Century Company, 1947.

\bibitem{yang2017measuring}
Ke~Yang and Julia Stoyanovich.
\newblock Measuring fairness in ranked outputs.
\newblock In {\em Proceedings of the 29th International Conference on
  Scientific and Statistical Database Management}, page~22. ACM, 2017.

\bibitem{young2001differential}
John~W Young.
\newblock Differential validity, differential prediction, and college admission
  testing: A comprehensive review and analysis. {R}esearch report no. 2001-6.
\newblock {\em College Entrance Examination Board}, 2001.

\bibitem{zafar2015fairness}
Muhammad~Bilal Zafar, Isabel Valera, Manuel~Gomez Rogriguez, and Krishna~P.
  Gummadi.
\newblock {Fairness Constraints: Mechanisms for Fair Classification}.
\newblock In {\em Proceedings of the 20th International Conference on
  Artificial Intelligence and Statistics}, volume~54, pages 962--970, Fort
  Lauderdale, FL, USA, 20--22 Apr 2017. PMLR.

\bibitem{zehlike2017fa}
Meike Zehlike, Francesco Bonchi, Carlos Castillo, Sara Hajian, Mohamed Megahed,
  and Ricardo Baeza-Yates.
\newblock {FA*IR}: A fair top-k ranking algorithm.
\newblock In {\em Proceedings of the 2017 ACM on Conference on Information and
  Knowledge Management}, pages 1569--1578. ACM, 2017.

\bibitem{zemel2013learning}
Rich Zemel, Yu~Wu, Kevin Swersky, Toni Pitassi, and Cynthia Dwork.
\newblock Learning fair representations.
\newblock In {\em International Conference on Machine Learning}, pages
  325--333, 2013.

\bibitem{zhou2015tackling}
Dawei Zhou, Jiebo Luo, Vincent~MB Silenzio, Yun Zhou, Jile Hu, Glenn Currier,
  and Henry Kautz.
\newblock Tackling mental health by integrating unobtrusive multimodal sensing.
\newblock In {\em Twenty-Ninth AAAI Conference on Artificial Intelligence},
  2015.

\end{thebibliography}

\clearpage

\appendix
\section{Administrative Information on Vendors}
\label{app:admin}

\begin{table}[ht]
  \centering
\begin{tabular}{llll}
\toprule
           Vendor name &       Funding & \# of employees &   Location \\
\midrule
           8 and Above &            -- &            1-10 &    WA, USA \\
              ActiView &        \$6.5M &           11-50 &     Israel \\
 Assessment Innovation &        \$1.3M &            1-10 &    NY, USA \\
              Good\&Co &       \$10.3M &          51-100 &    CA, USA \\
                Harver &         \$14M &          51-100 &    NY, USA \\
               HireVue &         \$93M &         251-500 &    UT, USA \\
            impress.ai &        \$1.4M &           11-50 &  Singapore \\
               Knockri &            -- &           11-50 &     Canada \\
                  Koru &       \$15.6M &           11-50 &    WA, USA \\
    LaunchPad Recruits &    \pounds 2M &           11-50 &         UK \\
           myInterview &        \$1.4M &            1-10 &  Australia \\
               Plum.io &        \$1.9M &           11-50 &     Canada \\
        PredictiveHire &       A\$4.3M &           11-50 &  Australia \\
             pymetrics &       \$56.6M &          51-100 &    NY, USA \\
             Scoutible &        \$6.5M &            1-10 &    CA, USA \\
             Teamscope &    \euro 800K &            1-10 &    Estonia \\
             ThriveMap &  \pounds 781K &            1-10 &         UK \\
                  Yobs &          \$1M &           11-50 &    CA, USA \\
\bottomrule
\end{tabular}

  \caption{Administrative information}
  \label{tab:admin}
\end{table}

\end{document}